# Highlights

## Privacy-Preserving for Images in Satellite Communications: A Comprehensive Review of Chaos-Based Encryption

Farrukh Bin Rashid,Windhya Rankothge,Somayeh Sadeghi,Hesamodin Mohammadian,Ali Ghorbani

- Provides a solid background of Satellite Communication and Image Encryption in Satellite Communication, including current industry standards.
- Analyzes and categorizes Chaos Based Image Encryption, considering the types of chaotic maps used.
- Delivers a comprehensive literature review on all state-of-the-art studies, specifically for Chaos Based Satellite Image Encryption.
- Presents a detailed analysis of the evaluation process for Chaos Based Image Encryption, including evaluation parameters and conditions.
- Discusses existing challenges and open research problems for Chaos Based Satellite Image Encryption, to facilitate future directions to upcoming research initiatives.

# Privacy-Preserving for Images in Satellite Communications: A Comprehensive Review of Chaos-Based Encryption ⋆,⋆⋆

Farrukh Bin Rashid[a,*,1], Windhya Rankothge[a], Somayeh Sadeghi[a], Hesamodin Mohammadian[a] and Ali Ghorbani[a]

[a]Canadian Institute for Cybersecurity, 46, Dineen Drive, Fredericton, New Brunswick, Canada
[a]Canadian Institute for Cybersecurity, 46, Dineen Drive, Fredericton, New Brunswick, Canada
[a]Canadian Institute for Cybersecurity, 46, Dineen Drive, Fredericton, New Brunswick, Canada
[a]Canadian Institute for Cybersecurity, 46, Dineen Drive, Fredericton, New Brunswick, Canada
[a]Canadian Institute for Cybersecurity, 46, Dineen Drive, Fredericton, New Brunswick, Canada

ABSTRACT

In an era where global connectivity has become critical, satellite communication is essential for businesses, governments, and individuals. Widely used services with satellite communication such as climate change monitoring, military surveillance and real-time event broadcasting, involve data in the form of images rather text. Therefore, securing image transmission in satellite communication using efficient and effective encryption approaches, has gained a significant attention from academia as well as the industry. In this paper, we specifically focus on chaos based image encryption as one of the key privacy-preserving techniques for satellite communication. While there are several privacy enhancing techniques for protecting image data but chaos based encryption has distinct advantages such as high flexibility, high security, less computational overheads, less computing power and ease of implementation. First, we present a solid background about satellite communication and image encryption in satellite communication, covering theoretical aspects of chaotic systems and their practical usage for image encryption. Next we present a comprehensive literature review on all state-of-the-art studies specifically for chaos based satellite image encryption, with a detailed analysis of the evaluation process, including evaluation parameters and conditions. Finally, we discuss about existing challenges and open research problems for chaos based satellite image encryption.

## 1. Introduction

Satellite systems have become a crucial component of global telecommunication systems, as they have been used in different types of communication systems such as broadband internet, weather forecasting, maritime communications, radio broadcasting, and etc. [1]. Satellite systems provide extensive coverage, providing communication from remote areas, disaster stricken regions and across borders. There are different types of Satellites, that can be distinguished by several factors, such as the shape of the orbit, travel direction and altitude of the satellites. The most popular way to distinguish satellites is by their altitude: Low Earth Orbit (LEO) satellites, Medium Earth Orbit (MEO) satellites and Geostationary Equatorial Orbit (GEO) satellites. LEO satellites are at an altitude of less than 1000 kms - closer to Earth with lower latency and less coverage. Therefore, LEO satellites are used for high-resolution earth observations, remote sensing, and scientific research as data can be acquired and transmitted rapidly. Furthermore, when deployed in constellations, LEO satellites also can be used for communication systems, as they can cover a larger area of the Earth [2]. MEO satellites are at an altitude of 5000 to 20000 kms, and extensively used for positioning and navigation services, such as Global Positioning System (GPS) and European Galileo System. GEO satellites are at an altitude of 36000 kms and remain in one place relative to the surface of the Earth, excelling at large-scale and long-term observations. Therefore, GEO satellites are used by TV and radio broadcast stations, weather forecasting services and etc. GEO constellations collect and transmit data for services such as weather, climate, and geological science [2]. Also, GEO satellites enable communication between the Earth and other orbiting objects, such as manned spacecraft and space telescopes.

*Corresponding author
farrukh.rashid@unb.ca (F.B. Rashid); windhya.rankothge@unb.ca (W. Rankothge); somayeh.sadeghi@unb.ca (S. Sadeghi); h.mohammadian@unb.ca (H. Mohammadian); ghorbani@unb.ca (A. Ghorbani)
ORCID(s):

In an era where global connectivity has become critical, satellite communication is essential for businesses, governments, and individuals. However, the rapid adoption of satellite communication in different applications has widened the threat surface of satellite systems, in-terms of physical security as well as cybersecurity. Satellite communications are susceptible to many cybersecurity vulnerabilities, such as hacking, malware infections, data breaches and supply chain attacks [3]. Viasat KA-SAT satellite network, which delivers high-speed internet services across Europe, was attacked in 2022 and it affected Ukrainian citizens heavily, with many experiencing internet outages lasting for two weeks [4]. The incident disrupted operations of a German energy firm responsible for the remote management of approximately 5,800 wind turbines, leading to a temporary loss of control and monitoring capabilities. Researchers believe that the attack was conducted by using a malware which was designed to wipe out vulnerable routers and modems. Furthermore, in March 2023, Starlink terminals in Ukraine were exposed to a signal jamming attack, causing a major disruption in services [5]. As a mitigation strategy, SpaceX had to implement software modifications.

With the evolving cyber threat surface, securing satellite communication systems has become challenging, specially the data transmission process. Incorporating robust encryption and authentication protocols is essential for securing satellite communication systems, where advanced encryption algorithms are expected to protect the confidentiality and integrity of data transmitted via satellites, while strong authentication mechanisms are expected to prevent unauthorized access. Widely used services with satellite communication such as climate change monitoring, military surveillance and real-time event broadcasting, involve data in the form of images rather text. Therefore, securing image transmission in satellite communication has gained a significant attention from academia as well as the industry, because most of the data that is transmitted in satellite communication systems are images. Images transmitted by satellites to ground control stations contain high-value, sensitive and proprietary information, making them prime targets for cyber attacks.

Image data contains set of unique features which are different from text data such as high redundancy of data, scattered information, large size data, strong correlation among adjacent pixels and bulk data capacity [6]. Therefore, even though the widely adopted method for satellite image encryption is Advanced Encryption Standard (AES), the nature of image data requires the exploration of encryption algorithms distinct from the conventional algorithms to ensure efficiency and effectiveness of the encryption process [7, 8]. During recent past years, there have been several research work carried out to explore secure image encryption algorithms such as Chaos based image encryption [9, 10], S-Box based encryption [11, 12], Optical encryption [13], Compression encryption [14], DNA-based image encryption [15] and Frequency domain based encryption [16].

In this paper, we specifically focus on chaos based image encryption as a privacy-enhancing technology for satellite communication. This approach is part of a set of privacy preserving techniques for images and videos, which includes various methods such as anonymization, obfuscation, and access control. Our focus is on chaos based encryption based on its advantages such as high flexibility, high security, less computational overheads, less computing power and ease of implementation. With the chaos based approach, the chaotic nature of mathematical algorithms to encrypt and decrypt images, ensure that even if the image being transmitted is intercepted, the images remain indecipherable without the correct key.

The key contributions of the paper can be summarized as follows:

- Providing a solid background of Satellite Communication and Image Encryption in Satellite Communication, including current industry standards.
- Analyzing and categorizing Chaos Based Image Encryption, considering the types of chaotic maps used.
- Delivering a comprehensive literature review on all state-of-the-art studies, specifically for Chaos Based Satellite Image Encryption.
- Presenting a detailed analysis of the evaluation process for Chaos Based Image Encryption, including evaluation parameters and conditions.
- Discussing existing challenges and open research problems for Chaos Based Satellite Image Encryption, to facilitate future directions to upcoming research initiatives.

There have been several surveys that elaborate on chaos based image encryption, its challenges and future directions. However, to the best of our knowledge, this is the first survey that provides a structured overview of chaos based image encryption, specifically focusing on Satellite images and Satellite communication. To accomplish this, first we provide a solid background in Section 2, about satellite communication and image encryption in satellite communication, including widely used recent technologies and current industry standards. Next in Section 3, we explore theoretical aspects of chaotic systems, directing our study into the practical usage of chaos based encryption approaches for image encryption.

Section 4 present a comprehensive literature review on all state-of-the-art studies, specifically for chaos based satellite image encryption. We have categorized the research work into following five different types, based on the types of chaotic maps used: (1) Logistic and Sine Maps, (2) Hyperchaotic and Multidimensional Systems, (3) Enhanced Chaotic Maps, (4) Integration of Multiple Chaotic Maps and (5) Chaotic Oscillators and Controllers. Each category of chaos based satellite image encryption, with the relevant existing research work that we explored are listed in Table 1. Furthermore, Table 2 also provides a comprehensive summary of each research work that we discussed, with further information such as the datasets utilized and the environments in which these approaches were evaluated. We Present a detailed analysis of the evaluation process for chaos based image encryption, including evaluation parameters and conditions in Section 5. Table 3, provide a summary of the evaluations performed by the research work that we explored, including different standard tests and cryptanalysis attacks. In 6, we discuss existing challenges and future research directions for chaos based satellite image encryption. We conclude our paper with the Section 7.

## 2. Encryption in Satellite Communication

With the recent advancements in satellite communication systems, satellites are used to provide services not only for traditional services such as broadband internet, weather forecasting and radio broadcasting, but also for wider range of emerging application domains, such as remote network connections with low bandwidth and low-power constrained Internet of Things (IoT) devices. Furthermore, most of these applications involve data in the form of images, and therefore, these days, most of the data that is transmitted in satellite communication systems are images. As a result, the threat surface of satellite systems has been widen, introducing many cybersecurity challenges, such as hacking attacks, malware infections, data breaches and supply chain attacks [3]. As countermeasures, robust end-to-end encryption and authentication protocols are implemented in satellite communication systems to ensure confidentiality and integrity of data transmitted, specially of the images, while preventing unauthorized access [17].

In this section we present an overview of satellite communication, and explore the encryption strategies in satellite communication, specifically for images. First, we discuss different industry standards and algorithms used for general encryption in satellite communication, including authenticated encryption approaches. Next, we specifically focus on image encryption, highlighting distinct requirements of image encryption in satellite communication, and providing examples of different image encryption approaches in the literature.

### 2.1. Satellite Communication Systems Architecture

Satellite networks generally span world-wide, and contain multiple subsystems, such as communication modules, power units, and navigation controls, making them vulnerable to a wide range of threats. Each component of a satellite communication system has its own security vulnerabilities, requiring specialized security measures. For the better understanding of security requirements, as shown in Figure 1, satellite communication systems are divided into three segments, where each segment requires a specific level of security to operate effectively and securely.

- Space Segment: Comprises of GEO, MEO and LEO satellites including Satellite to Satellite (SS) links and Satellite to Ground (SG) links. Povides services such as navigation, data, mobile television and radio broadcasting systems.
- Ground Segment: Comprises of Ground Controllers and Network Controllers including Ground to Ground (GG) links and Ground to Satellite (GS) links. Enables communication between the ground and satellites.
- User Segment: Comprises of user terminals such as smartphones, airplanes, ships and etc. Communicates with the satellite and ground segments using forward links.

### 2.2. Satellite Communication Encryption

Given the increasing number of cyber threats targeting satellite communication systems, it is crucial to deploy robust security measures, that address existing vulnerabilities as well as anticipated future threats. Industry and academia have shown a keen interest on exploring and implementing cost effective security strategies through standardisation, specially the Consultative Committee for Space Data Systems (CCSDS). The CCSDS was formed in 1982 by major space agencies to provide a platform to discuss problems in the development and operation of space data systems. They have published several resources focusing on cybersecurity aspects of satellite communication systems such as CCSDS 350.1-G-2 Security Threats Against Space Missions, CCSDS 350.7-G-1 Security Guide for Mission Planners, CCSDS 352.0-B-1 CCSDS Cryptographic Algorithms and CCSDS 355.0-R-3 Space Data Link Security Protocol [18]. The CCSDS highlights three security requirements for satellite communication systems: (1) protection of information sent from the ground station to a satellite to command or control its functions, (2) protection of information transmitted from the satellite back to the ground station, and (3) protection of data within the ground station [19, 20].

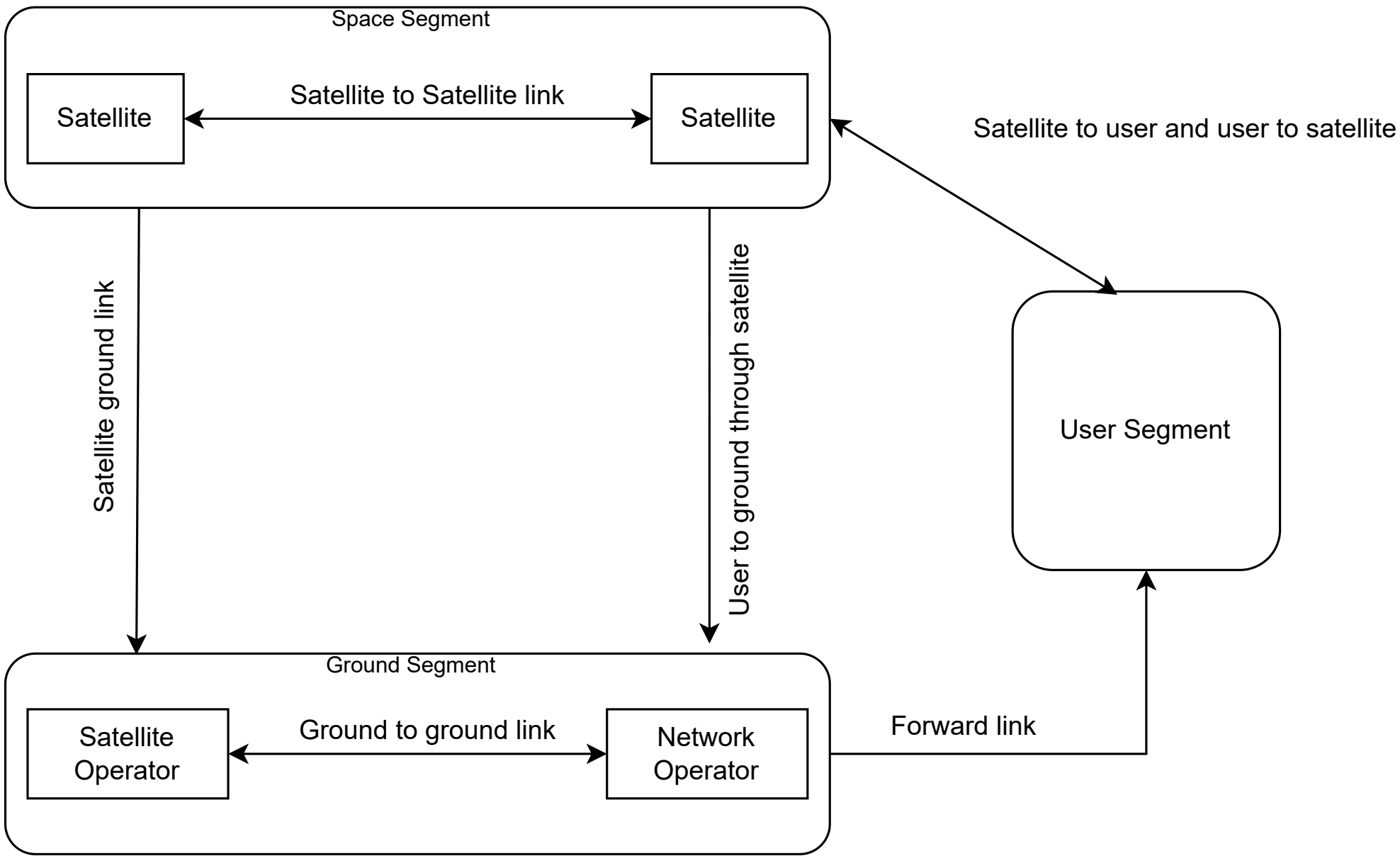

**Figure 1:** Satellite Communication Systems Architecture

Following the recommendations and standards of CCSDS, incorporating a robust encryption mechanism is essential for securing satellite communication systems, where advanced encryption algorithms preserve the confidentiality and integrity of data transmitted via satellites. Encryption algorithms that are being used in modern communication systems can be categorized into two types:

- Symmetric Encryption: Employs the same key for both encryption and decryption. Since a single key is being used, encryption is fast, but sharing the key between communicating parties securely is challenging. Examples include Advanced Encryption Standard (AES), Data Encryption Standard (DES), Triple Data Encryption Standard (3DES) and Blowfish
- Asymmetric Encryption: Employs a pair of keys: a public key for encryption and a private key for decryption. resolves the key distribution challenge, but is comparatively slower. Examples include Rivest-Shamir-Adleman (RSA), Diffie-Hellman and Elliptic Curve Cryptography (ECC).

The CCSDS recommends Symmetric Encryption algorithms for satellite communication systems where delays are critical and computational resources are limited [21]. In general implementation of symmetric encryption, first a key-exchange protocol is used for establishing a shared key between the entities, and then this key is further used to authenticate and encrypt the transmitted data using a symmetric-key algorithm. However, since the wireless medium is used for authentication, it is prone to impersonation attacks [17]. Therefore, several methods have been proposed over the years to securely share keys between parties involved, and then to implement authentication protocols. Elliptic Curve Cryptography (ECC) is one of the proposed security approach to authenticate peer entities [22, 23, 24]. ECC operates on the principle of elliptic curves algebraic structure within finite fields. The security of ECC relies on the idea that cracking the Elliptic Curve Discrete Logarithm Problem (ECDLP) is computationally infeasible. Furthermore, research work on [25, 26, 27, 28, 29] propose a Pre-Shared key authentication approach, which refers to a secret key that is shared in advance between communicating entities. When one entity initiates communication with the other, they present the Pre-Shared key as part of the authentication process. The authors of [30, 31, 32, 33] employ the Timed Efficient Stream Loss Tolerant Authentication (TESLA) protocol, which is designed to ensure delayed source authentication in broadcast communications, which adds a layer of security while accommodating the challenges of broadcast communication. TESLA protocol uses symmetric cryptography primitives and hash chains contributes to its efficiency, making it suitable for resource-limited environments. All these proposed schemes have the security property of authenticating before initiating two-way communication and performing message authentication, in order to ensure that the message truly comes from the entity it claims to be from [17].

Furthermore, a key establishment protocol can be used to facilitate communicating entities to reach an agreement on a shared key, which then helps satellites in further mutual communications [34, 17]. Most of the techniques proposed for key sharing can be used for the implementations of key agreement protocols as well, focusing on providing confidentiality, integrity, availability and also identity verification at the same time [35, 17]. Identity-Based Encryption (IBE) scheme is one of the approaches, where the ground control and the satellite employ distinct identifiers to calculate their respective public keys. In the IBE system, no advance enrollment or certificates are necessary. However, the scheme includes a key generation process which is vulnerable to potential key escrow attacks [36, 17]. Another approach is the use of chaotic maps, which are faster, since there is no modular arithmetic and are also computationally efficient, since large keys are not required. However, the cipher text produced by chaotic maps can be lengthier than the plain text.

## 2.3. Authenticated Encryption

As we discussed in the previous section, securing satellite communication essentially focuses on preserving three fundamental cybersecurity features: confidentiality, integrity and authenticity, which can be achieved through robust encryption and authentication mechanisms. Traditional encryption algorithms are designed to ensure confidentiality, whereas authentication algorithms focus on providing authenticity and integrity services [37]. However, Authenticated Encryption algorithms encompass all three features, guaranteeing data security through encryption and adding verification measures to ensure data integrity.

The CCSDS recommends Symmetric Encryption algorithms for satellite communication systems where delays are critical and computational resources are limited [21]. Therefore, Authenticated Encryption algorithms that employ Symmetric Encryption are preferred for satellite communication systems to preserve confidentiality, integrity and authenticity of the data that is being transferred. The widely adopted method for satellite image encryption is Advanced Encryption Standard (AES). Also, it is important to note that the use of 3DES is deprecated for all applications through 2023 and disallowed after December 31, 2023, due to security weaknesses resulting from its small key size, according to National Institute of Standards and Technology (NIST) [38].

Another critical aspect of satellite communication systems is to ensure high-speed communication between ground control and satellites, which requires fast and efficient encryption algorithms.

Most widely used approach for fast and efficient encryption is AES - Cipher block Chaining Message authentication code (AES-CCM), which combines AES - Counter (AES-CTR) with the Cipher Block Chaining Message Authentication Code (CBC-MAC) [39]. AES-CTR represents a mode of operation for AES, transforming it from a block cipher, which encrypts fixed-size blocks, into a stream cipher mode. This transformation is achieved through the use of an Initialization Vector (IV), which is segmented into three components: a nonce, a constant vector, and a counter vector. Then a key stream is generated, which is a sequence of pseudo-random bits. This key stream is then XOR-ed with the plain text block to produce the cipher text block [40, 37]. However, AES-CCM is generally not recommended for satellite systems, as satellite systems require high data throughput.

The GCM (Galois Counter Mode) algorithm is endorsed by the CCSDS for use in satellite applications [21], which gives a higher data throughput [41]. The GCM is an authenticated encryption algorithm with associated data mode, where it uses a counter mode for encryption but adds a Galois field multiplication for authentication. An initialization vector is used in GCM, too, which should be unique for each message to ensure security [37].

When AES-CTR and GCM algorithms are combined together, the AES-CTR is used as the encryption algorithm to generate the cipher text blocks, while the GCM is sued to provide. authentication and integrity for the cipher text generated by AES-CTR. The AES-CTR and GCM algorithms employ distinct, unique symmetric keys to generate ciphertext and authentication tags. However, to enhance security and mitigate the risk of attacks, particularly in security-sensitive domains such as satellite communications, it is generally recommended not to use the same key for symmetric encryption for a prolonged period. Therefore, Perfect Forward Secrecy (PFS) approch is employed, which generates a new key for every session. With PFS, if a single key is compromised, only the session or message associated with that key is at risk, while the rest of the messages or sessions stay protected. [20].

## 2.4. Image Encryption in Satellite Communication

Satellite communication facilitates wide range of services and applications that process data not only in the form of text, but also in the form of images. Most of the data that is transmitted in recent satellite communication systems are images. As these images can be confidential and highly sensitive data, it is required to employ secure encryption algorithms. When focusing on satellite image transmission, generally, traditional encryption algorithms such as 3DES, DES, IDEA, and RSA are used as encryption algorithms. However, DES and 3DES are less secure than AES due to their key sizes [42] and therefore, widely adopted method for satellite image encryption is AES, which is also endorsed by the CCSDS.

Image data contains set of unique features which are different from text data such as high redundancy of data, scattered information, large size data, strong correlation among adjacent pixels and bulk data capacity [6]. Large image encryption may lead to a slower encryption process and higher resource utilization [42]. The image encryption process generally follows 2 steps: (1) convert the two-dimensional (2D) image into a one-dimensional (1D) data stream and (2) encryption using text based encryption. In the text based encryption step, the decrypted text should always be identical to the original text, because to decrypt the original message, each bit must be recovered very precisely. However, for digital multimedia applications, there is no such requirement because, a minute change in the pixel attribute of an image, will not degrade the image quality drastically [6]. Furthermore, pixel data of an image consists intensity values in the range of [0, 255]. If we use a conventional algorithm to convert images, the encrypted value for a pixel is set as per the encryption key used. Since the pixel value repeats in an image multiple times, the value could be guessed by the attacker easily [6]. Therefore, even though the widely adopted method for satellite image encryption is AES, the nature of image data requires the exploration of encryption algorithms distinct from the conventional algorithms to ensure efficiency and effectiveness of the encryption process [7, 8].

When designing and implementing image encryption algorithms, it is important to note following unique characteristics of images:

- Large Capacity: Images hold substantial information requiring an encryption algorithm that considers the extensive amount of data, ensuring efficient encryption and decryption process.
- High Redundancy: An image transmitted by a satellite may contain repetitive or duplicated information. Redundancy in images can manifest as similar patterns, textures, or color information occurring in multiple areas of the image. As an example, in an image of a blue sky with white clouds; it exhibits a high level of redundancy as the blue color and cloud patterns often repeat across different sky regions. The encryption algorithm must effectively obscure or randomize such redundant patterns, because if the algorithm neglects this characteristic, it could expose the encrypted data to specific attacks such as plain text attacks and differential attacks. In plain text attacks, if the attacker gets access to the plain and the encrypted images, he will be able to analyze the similarities and deduce the key or encryption algorithm being used. In differential attacks, attacker can take near identical images and compare their encrypted versions, which will allow the attacker to find patterns, which in turn allows him to understand the encryption process.
- High Correlation between Pixels: Refers to the statistical relationship or dependence among the values of neighboring pixels within an image. When the color/intensity of one pixel is related to the color/intensity of its neighboring pixels, there is a high correlation. If the encryption algorithm does not adequately disrupt or randomize the correlation between pixels, it might be exposed to specific attacks. As an example, a frequency analysis attack can exploit consistent patterns in the encrypted data. Therefore, the encryption algorithm should introduce sufficient complexity and randomness to obscure these patterns.

Considering the above unique characteristics, image encryption algorithms should be designed to ensure the confidentiality, integrity, and privacy of sensitive visual data. Furthermore, image encryption algorithms can also be applied to digital watermarks and copyright information to preserve them. However, it is essential to note that while image encryption algorithms help in mitigating security threats, they do not offer complete protection against breaches or tampering [7].

During recent past years, there have been several research work carried out to explore secure image encryption algorithms such as Chaos based image encryption [9, 10], S-Box based encryption [11, 12], Optical encryption [13], Compression encryption [14], DNA-based image encryption [15] and Frequency domain based encryption [16].

In this paper, we specifically focus on chaos based image encryption, and its usage for satellite communication, because of the advantages provided by chaos based image encryption, such as high flexibility, high security, less computational overheads, less computing power and ease of implementation.

## 3. Chaos Based Image Encryption

As we mentioned in the previous section, the unique features of image data require the exploration of new encryption algorithms, and chaos based image encryption algorithms have been proved to be efficient and effective. Chaotic systems are described as a set of dynamical equations that vary with time, and time can be discrete or continuous. It is a predictable dynamic system that exhibits seemingly random behaviors, attributed to its extreme sensitivity to initial conditions and parameters. The roots of chaos theory can be traced back to the solution of the three-body problem in 1913 [43]. Nonetheless, the first instance of a chaotic solution stemming from a deterministic system's equation was identified by the work in [44]. Furthermore, in 1976, the Logistic map, which is one of the widely used chaotic map, was proposed in the work [45]. Since then the study of chaotic systems has progressed exponentially, and chaotic systems have been used in wide range of application domains. The distinct features of chaotic systems such as determinacy, ergodicity and sensitivity to initial condition, have make them a better choice for designing secure encryption algorithms.

In this section we present an overview of chaotic systems, and explore chaos based image encryption. First, we discuss different chaotic systems, including Discrete chaotic and Continuous chaotic. Next, we discuss how chaotic systems are being used for image encryption, highlighting the benefits of chaos based image encryption.

### 3.1. Chaos Based Systems

As shown in Figure 2, Chaotic systems fall into two main categories: (1) Discrete chaotic and (2) Continuous chaotic [46], based on the system's evolution. While the system's evolution occurs in discrete time intervals in discrete chaotic systems, the time is considered as a continuous variable in continuous chaotic systems, and the system's evolution occurs smoothly.

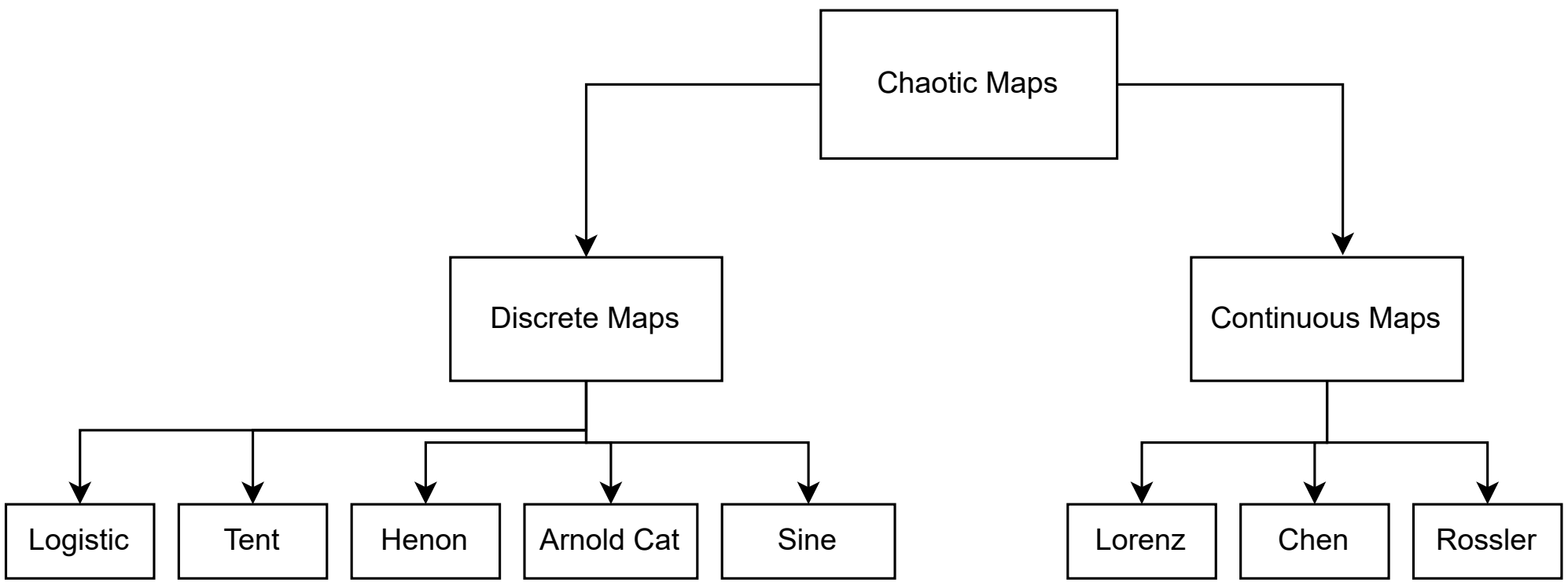

**Figure 2:** Categorization of Chaotic Systems

#### *3.1.1. Discrete Chaotic Systems*

The discrete chaotic systems, employ distinct features such as being non-linear, sensitive to initial conditions, and period multiplication. Discrete chaotic maps can be used in various applications, such as digital signal processing, communication system modulators, image encryption and compression [47]. Following are examples of fundamental discrete chaotic systems:

- Logistic Map [48, 49]: A mathematical equation often used as an example of a discrete dynamical system, that exhibits chaotic behavior under certain conditions. Here, the $X_t$ is a state variable where the parameter $p$ is crucial in determining the behavior of the chaotic map. $p$ must be between 0 and 4 to exhibit exciting dynamics. The chaotic map exhibits complex, chaotic behavior when between 3.74 and 4.

$$X_{t+1} = f_l(X_t) = pX_t(1 - X_t) \tag{1}$$

- Tent Map [50]: Operates in one dimension where it exhibits chaotic behavior under certain conditions. The value of $p$, a control parameter, is between 0 and 2, which is required for the system to exhibit interesting dynamics. The tent map is piece-wise linear with a tent-shaped form.

$$X_{t+1} = f_t(X_t) = \begin{cases} pX_t, & \text{if } X_t < \frac{1}{2}, \\ p(1 - X_t), & \text{if } X_t \geq \frac{1}{2}, \end{cases} \tag{2}$$

- Henon Map [51]: A two-dimensional, non-linear, and dissipative system. The $X$ and $Y$ represent a point in a two-dimensional space at time $t$, and $a$ and $b$ are parameters that influence the map's behavior.

$$\begin{cases} X_{t+1} &= 1 - aX_t^2 + Y_t, \\ Y_{t+1} &= bX_t, \end{cases} \tag{3}$$

- Arnold Cat Map [52, 53]: A two-dimensional chaotic map, that operates on the coordinates of a point in a square grid. It is known for periodic behavior, repeating after a certain number of iterations. Despite the periodicity, the map is considered chaotic due to its dependence on the initial conditions.

$$\begin{cases} X_{t+1} = (2X_t + Y_t) \bmod 1, \\ Y_{t+1} = (X_t + Y_t) \bmod 1, \end{cases} \tag{4}$$

- Sine Map [54]: A one-dimensional dynamical system, known for its chaotic behavior within specific parameter ranges. The parameter $\mu$ is usually kept between 0 and 2 to induce interesting and chaotic dynamics. The non-linearity introduced in the sine map is due to the sine function, resulting in a piece-wise continuous and oscillatory behavior reminiscent of the sine curve.

$$\begin{cases} X_{t+1} = \mu \sin(\pi x_t) \end{cases} \tag{5}$$

### *3.1.2. Continuous Chaotic Systems*

The continuous chaotic systems, employ distinct features such as displaying intricate and unpredictable behavior over time. These systems are characterized by a collection of ordinary or partial differential equations, that prescribe the alterations in state variables as they progress through time. These state variables denote measurable physical characteristics inherent to the scrutinized system, such as pressure, temperature, position, or velocity. Following are examples of fundamental continuous chaotic systems:

- Lorenz Equation [45]: Describes the rate of change of three quantities concerning time: $x$ (proportional to the rate of convection), $y$ (horizontal temperature variation), and $z$ (vertical temperature variation). The other parameters of the system are constants. The Lorenz system stands as a highly examined instance of chaos arising from deterministic, non-linear dynamic equations.

$$\begin{cases} \frac{dx}{dt} = \sigma(y - x), \\ \frac{dy}{dt} = \rho x - y - xz, \\ \frac{dz}{dt} = xy - \beta z, \end{cases} \tag{6}$$

- Chen System [55, 56]: An example of a chaotic system that exhibits rich, dynamic behavior with a simple structure relatively. The parameters $a$, $b$ and $c$ govern the system's behavior, and slight parameter variations can lead to vastly diverse and different dynamics.

$$\begin{cases} \frac{dx}{dt} = a(y - x), \\ \frac{dy}{dt} = (c - a)x - xz + cy, \\ \frac{dz}{dt} = xy - bz, \end{cases} \tag{7}$$

- Rossler System [57, 58]: Was initially devised as a mathematical model for chemical reactions, and known for its spiral nature.

$$\begin{cases} \frac{dx}{dt} = -(y + z), \\ \frac{dy}{dt} = x + ay, \\ \frac{dz}{dt} = b + z(x - c), \end{cases} \tag{8}$$

### 3.2. Chaos Based Image Encryption

Due to previously mentioned unpredictable and complex behavior of chaotic systems, they can be applied to a wide range of application domains. In finance, chaotic systems model stock market behavior, allowing traders to develop different trading strategies. In machine learning, chaos theory is used to model the neurons' behavior, allowing researchers to develop new algorithms. They are also used in biology to study the behavior of biological systems. Chaotic patterns are also found in a variety of natural systems, such as the flow of fluids, climatic variations and irregular heartbeats. Furthermore, cryptography and encryption often uses chaotic systems to create secure communication systems, leveraging their inherent structural similarities [59, 60, 10].

Delving more into chaos based encryption, one of the initial digital image encryption algorithms was proposed in 1991, with the use of chaos discrete models [10]. As shown in Figure 3, the general chaos based image encryption process follows of two phases: (1) the Confusion Phase and (2) the Diffusion Phase.

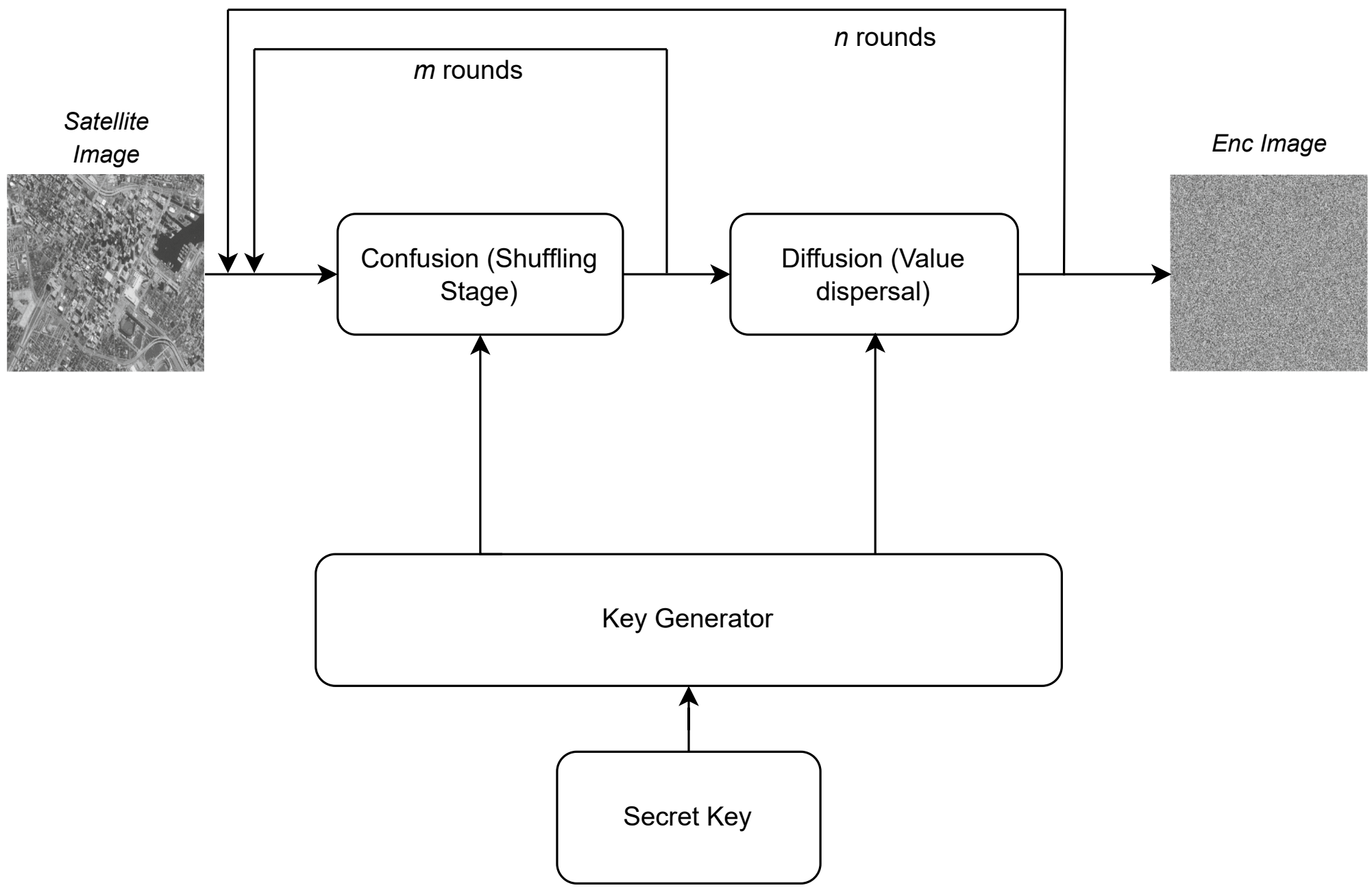

**Figure 3:** General Chaos Based Image Encryption Process

Initially, the plain image is passed to the Confusion Phase which is also known as the pixel permutation phase. In the Confusion Phase, the position of the pixels are changed, but the pixel values are kept unchanged, which converts the image into an unidentifiable format. Next, the Diffusion Phase is executed with the use of a chaotic map, to tighten the security of the encryption process. In the Diffusion Phase, the pixel values are sequentially modified by the sequence generated from the chaotic system, so that even a small change in the pixel, affects majority of the pixels in the entire image. The Confusion–Diffusion processes are iteratively performed for multiple times, until a satisfactory level of security is obtained.

Chaotic systems are more suitable for encrypting images, due to following inherent characteristics:

- High Ergodicity: A chaotic system has a strong tendency to thoroughly explore and cover its entire state space. The system has complex and unpredictable behavior, visiting various states, and does not get stuck in certain regions of the state space. Instead it explores various states, making it more difficult to predict or analyze. In image encryption, having a chaotic system with high ergodicity property is an advantage, as the system will effectively blends and rearranges pixels within an image, making it challenging for an attacker to discern any recognizable patterns or structures in the encrypted image.
- Pseudo-Randomness: Pseudo-Randomness refers to the generation of random sequences of numbers or patterns that are not truly random, but exhibit characteristics of randomness. Chaotic system's Pseudo-Randomness is useful for image encryption algorithms, since they can generate pseudo-random sequences or keys to encrypt pixels within the image.
- Determinism: Determinism refers to the predictability of a system's future given its current state. Chaotic systems are highly sensitive to small changes in initial values and their evolution is entirely determined by initial conditions and system parameters. Chaotic systems generate seemingly random sequences through deterministic rules. Deterministic chaos introduces complexity, making it challenging for attackers to decipher encrypted images, which enhances encryption security.

## 4. Chaos Based Image Encryption in Satellite Communication

Most of the recent applications of satellite communications such as weather forecasting, internet and TV broadband, IoT networks, transmit data in the form of images. Image data have inherent characteristics such as large capacity, high redundancy and high correlation between Pixels. Considering the above unique characteristics, image encryption algorithms are designed to ensure the confidentiality, integrity, and privacy of sensitive visual data. As we mentioned in previous sections, chaos based encryption algorithms have been widely used in satellite image encryption, due to several reasons such as their ideal cryptographic properties, initial value sensitivity, and pseudo-randomness.

In this section, we comprehensively discuss several research work that have explored chaos based image encryption approaches, specifically for satellite communication, with information such as the proposed chaos based encryption process, datasets utilized and the environments in which these approaches were evaluated.

We have categorized the research work into following five different types, based on the types of chaotic maps used: (1) Logistic and Sine Maps, (2) Hyperchaotic and Multidimensional Systems, (3) Enhanced Chaotic Maps, (4) Integration of Multiple Chaotic Maps and (5) Chaotic Oscillators and Controllers.

Each category of chaos based satellite image encryption, with the relevant existing research work that we explored are listed in Table 1. Furthermore, Table 2 also provides a comprehensive summary of each research work that we discussed, with further information such as the datasets utilized and the environments in which these approaches were evaluated. Additionally, it is important to note that the majority of the referenced research work evaluated their proposed encryption approaches using different image datasets, and therefore, it is difficult to compare the encryption approaches directly.

### 4.1. Logistic and Sine Maps

The first category we explored was the research work that employs Logistic and Sine maps for chaos based encryption. Logistic and Sine maps are foundational discrete maps discussed in Equation (3.1.1) and (5). There are several research work that employs Logistic and Sine maps, individually or together to implement secure and robust satellite image encryption algorithms.

Authors of [61] propose a chaotic encryption algorithm for satellite images, based on 2D filtering and fisher yates shuffling that employs a combination of logistic and sine maps, to enhance security and reduce encryption time. First, the image is converted into a vector, divided into several blocks of fixed-size vectors, and then transformed into a matrix. Next, a filtering operation is generated using a 1D sine-powered chaotic map [80], which is applied to the transformed matrix, adding diffusion to the cryptosystem. To add confusion to the algorithm, the rows and columns of the filtered and transformed matrix are scrambled using the modern and classical Fisher-Yates shuffling [81]. Fisher-Yates shuffling generates a random permutation of a finite set. The two Fisher-Yates shuffling are obtained using 2D hyper-chaotic maps, the 2D logistic ICMIC cascade map [82], and the 2D-SLIM chaotic map [83]. The implementation was evaluated on three different satellite images: River image (1024×768), Denver image (800×600), and Tokyo image (2596×1920). As the key space of an encryption algorithm should be more than $2^{100}$ [84] to avoid being prone to brute-force attacks, the proposed algorithm uses secret keys, initial condition values for all three chaotic maps, and has a keyspace equivalent to $2^{250}$. Furthermore, within the data loss attack analysis, the image quality dropped when (128x128) blocks are removed from the image, while the PSNR was >23 when the block size is (64x64) and (32x32). The authors also compared the encryption time with the classic AES encryption time, where the proposed algorithm was 20% faster on the (1025x768) river image and 11% faster on the (2596x1920) Tokyo image.

**Table 1**
Summary of different types of satellite image encryption research, based on the types of chaotic maps used

| Category | Type of satellite image encryption research | Existing Work |
|---|---|---|
| 1 | Logistic and Sine Maps | [61, 62, 63, 64] |
| 2 | Hyperchaotic and Multidimensional Systems | [42, 65, 66, 67, 68] |
| 3 | Enhanced Chaotic Maps | [69, 70, 71, 72] |
| 4 | Integration of Multiple Chaotic Maps | [73, 74] |
| 5 | Chaotic Oscillators and Controllers | [75, 76] |

**Table 2**
Existing literature for Satellite Image Encryption Algorithms

| Research | Category | Dataset | Environment | Encryption time(s) / Image size |
|---|---|---|---|---|
| [61] 2023 | 1 | Image of a river (1024×768), Image of Denver (800×600), Image of Tokyo (2596×1920) | Pentium I3, CPU 2.2 GHz, Windows 7, 2GB RAM, Matlab V. 2017a | 0.748 - Tokyo image (2596x1920) |
| [62] 2021 | 1 | CompleteSun, Sun, Star, Cosmic Fair Lights images | Intel Core I3, Windows 10, 6GB RAM | NA |
| [63] 2019 | 1 | Gray scale cameraman image (256×256) | MATLAB 7 | 0.8211 - Lena Image (256x256) |
| [64] 2017 | 1 | Gray-scale images of Mont Saint Michel (2048x2048) | Intel Core I5, CPU 1.8GHz, 6GB RAM, Matlab | 2.818 - Mont Saint Michael image (2048x2048) |
| [42] 2020 | 2 | 32-meter resolution image of Oran, 2.5-meter resolution image of Algiers, 4-meter multi-spectral resolution image of Vancouver | Pentium I7, CPU 2.3 GHz, Windows 7, 4GB RAM, Matlab | 32.65s - Oran image (3x10000x12300) |
| [65] 2020 | 2 | Vancouver Image (1311x873) with 4 spectral bands: near-infrared, red, green and blue | NA | NA |
| [66] 2023 | 2 | Port image (720x358), Modern City image (1024x614), Florida image (2196x2196) [77] | Pentium I3, CPU 2.2GHz, Windows 7, 2GB RAM | 0.4923 - Florida Image (2196x2196) |
| [67] 2021 | 2 | Image of Aleppo City (1563x117), Image od Port (1500x999), Uber image (730x519) | Pentium I3, CPU 2.2GHz, Windows 7, 2GB RAM, Matlab V. 2017a | 0.120087 - Aleppo City image (1563x117) |
| [68] 2021 | 2 | Images from Signal and Image Processing Institute (SIPI) database [78] | NA | NA |
| [69] 2021 | 3 | sat1, sat2, sat3 (512x512) satellite images | Intel Core I7, CPU 2.9GHz, Windows 10, 16GB RAM, Mathematica 11 for key generation and MATLAB R2018b | 0.3137 - Random Image (1024x1024) |
| [70] 2023 | 3 | 'Kyiv' Image (512×512), 'San Diego' Image (512x512) | MatlabR2020b is used | 0.983 - Lena Image (256x256) |
| [71] 2018 | 3 | Singular image, size or type not mentioned | Intel Core I5, CPU 2.4 GHz, 4GB RAM, Microsoft C#.net V6, visual studio 2017 | NA |
| [72] 2020 | 3 | Tripoli, Rio De Janeiro, Stockholm images (10 bits) of WorldView-3 satellite [79]. Size and resolution of images are not specified. | NA | NA |
| [73] 2018 | 4 | Satellite images from United States Geological Survey (USGS). | Intel Core I7, CPU 3.6GHz, 16GB RAM. | NA |
| [74] 2019 | 4 | NA | NA | NA |
| [75] 2021 | 5 | Boston image (1400×700) | NA | NA |
| [76] 2021 | 5 | Ten images (256x256) from Gaofen Image Dataset (GID) | Intel Core I7, CPU 2.5GHz, Windows 10, 8GB RAM | 0.2293 - Average of 10 GID images (256x256) |

Research work on [62] used logistic maps during the key generation process in satellite image encryption, for creating a pseudo-random key. The key is subsequently integrated with the Logistic map (LM) [85], with the Cosine transformed Logistic map (CTLM), and with the Cosine transformed Logistic-Sine Map (CTLSM) [86, 87], leading to the development of three distinct image encryption algorithms. All the algorithms used the pseudo-random 384-bit initial shared key, which indicates a huge key space. First, the shared key is segmented into three sections, each consisting of 128 bits tailored for the Red, Green and Blue (RGB) channels of the plain image. Next, these keys are further split into four segments, each comprising 32 bits, and resulting in three pseudo-random keys designated for each component of the image. Following the generation of the pseudo-random key, a chaotic map is generated. All three, the plain image, chaotic map, and the pseudo-random key, are used to generate the cipher image. The combinations of each chaotic map with the pseudo-random key are then analyzed by testing on parameters such as Correlation coefficient, Entropy, Number of Changing Pixel Rates, and etc. The dataset used by the study consists of the "CompleteSun", "Sun", "Star" and "Cosmic Fair lights" images, taken by the "NASA Goddard Space Center". The proposed scheme took around 8 seconds to encrypt/decrypt a (512x512) image with a low performances system. However, authors have not done a comparison with the classical AES encryption algorithm or other existing research work. Furthermore, the paper does not consider the transmission errors and Single Event Upsets that might occur during the transmission of the images.

Authors of [63] introduced an image encryption algorithm inspired by biological systems and based on chaotic maps. Initially, the algorithm utilizes the logistic map to generate four chaotic sequences, and then the image goes through a series of operations. First is the crossover operation, which is responsible for changing the order of image pixels both row-wise and column-wise, representing the confusion part of the encryption. Second is the mutation operation, which takes the image obtained from the crossover step and XORs it with a randomly selected pixel from a secret random image; the random pixel selection is based on the chaotic sequences. Finally, the scrambler operation modifies the pixel values' locations in the altered image through a random sequence. Furthermore, if the image is RGB, the three operations are performed separately for each color channel. The algorithm is evaluated by calculating correlation coefficients, information entropy and performing differential attack on the encrypted images.

The algorithm proposed in [64] used a chaos model based on three different chaotic maps (chaotic maps, logistic chaotic generator, and sine chaotic generator) and the graph theory rotation to encrypt satellite images. One of the three different chaotic maps are selected by using the Standard Chaotic Generator, and chaotic keys are generated from the selected map. Next, the original image is XORed with the original image. Furthermore, cyclic rotations based on graph theory have been used to improve image scrambling and to strengthen the resistance against plain image attacks. The proposed algorithm uses a key space of up to $2^{512}$, which is desirable to resist brute force attacks. The study employed SPOT images (gray-scale images of Mont Saint, Michel, France) of (2048x2048) for the evaluation. The research also determined the maximum deviation, quantifying the extent of difference between the original and encrypted images, where a higher deviation value indicates a greater divergence of the encrypted image from the original image [88].

## 4.2. Hyperchaotic and Multidimensional Systems

The second category we explored was the research work that employs hyperchaotic and multidimensional systems for chaos based encryption. Hyperchaotic maps are dynamic systems, characterized by a higher number of positive Lyapunov exponents than traditional chaotic maps. Lyapunov exponents are mathematical quantities used to describe the rate of divergence or convergence of nearby trajectories in a dynamic system. Given that a hyperchaotic system possesses multiple positive Lyapunov exponents, it exhibits a greater level of complexity and unpredictability [89]. Furthermore, multidimensional chaotic maps exhibit chaotic behavior in more than one dimension [64]. Therefore, several studies have been conducted to explore the use of hyperchaotic and multidimensional systems for satellite image encryption approaches.

Authors of [66] introduced a framework that integrates a 7D hyperchaotic system, a pseudo-AES algorithm, and a permutation scheme, for satellite image encryption. The 7D hyperchaotic system was used in four different operations of the proposed encryption algorithm. The 7D hyperchaotic was preferred over other chaotic systems because the first Lyapunov exponents are greater than zero, and it exhibited a more intricate dynamic behavior, with phase trajectories diverging in multiple directions [90]. The initial key for the pseudo-AES algorithm was generated using the first two variables of the 7D system. Moreover, the pseudo-AES algorithm included multiple rounds of processing, where the first and last rounds were predetermined, following the traditional AES algorithm. In contrast, the remaining rounds in each block are determined by the hyperchaotic system's third variable (z). The plaintext image was fed as an input to the pseudo-AES algorithm. The output of the pseudo-AES was then subjected to a permutation, which used the fourth and fifth variables of the 7D hyperchaotic system to scramble the image further. Finally, the last variables of the 7D system were used to generate a mask matrix, which was then XORed with the previous step, obtaining the final image. Since their initial values determined the secret key for chaotic algorithms, the 7D chaotic system boasted a keyspace of $2^{350}$, making it the algorithm with one of the most extensive keyspaces in the satellite image encryption domain. This vast keyspace contributed significantly to the algorithm's robust security against brute-force attacks. The evaluation was performed on three different satellite images: Florida Image (2196x2196), Modern City Image (1024x614), and Port Image (720x358). The data loss attack was exclusively executed on the (720x358) Port image, deteriorating the deciphered image quality to below 20dB (PSNR), when the data loss is increased to (128x128). When comparing with traditional AES, the (2196x2196) image was encrypted in 1.3425 seconds by AES, while the proposed algorithm only took 0.4923 seconds, showing that the proposed algorithm performed well even on high-resolution images.

The work in [67] developed an innovative encryption algorithm for satellite images that leverages the Linear Feedback Shift Register (LFSR) generator, SHA-512 hash function, hyperchaotic systems, and the Josephus problem [91]. The algorithm uses two hyperchaotic systems: the 1D hyperchaotic Logistic-tent system [92] and the 6-D hyperchaotic system [93]. For the encryption process, a Logistic Tent 1-D hyperchaotic system was employed as a control mechanism, initially selecting equations of the 6-D hyperchaotic system based on the controller's output, then XORing the selected equations with the original image, next applying the scrambling operations generated by Josephus sequences, and finally repeating these steps for a specified number of rounds to produce the final cipher image. The hyperchaotic system improved the diffusion process in the algorithm, while the Josephus problem [94] elevated the security measures of the cryptosystem during the confusion process. The keyspace of the algorithm was $2^{326}$, which is large enough to resist any brute force attacks. The evaluation of the algorithm was conducted using three distinct satellite images: an image of Aleppo city (1563x117), a Port image at (1500x999), and an Uber image (730x519). The algorithm performed reasonably well against data loss attacks even after cropping a (128x128) data block. However, the time complexity for encrypting a traditional (256x256) 'Lena' image [95] using this algorithm was notably higher, requiring approximately 7 seconds.

Authors of [68] proposed a hybrid encryption scheme that combines chaotic dynamical systems, specifically three-dimensional Henon and Chen chaotic maps, with the Blowfish encryption algorithm [96]. A three-layer digital image served as the starting point in this encryption process, where Two three-dimensional chaotic dynamical systems were iterated to generate Pseudo-Random Number Generators (PRNGs), and then underwent bitwise XOR operations. The transformed images were converted into 1×n direction layers. The criteria for word length collection were defined to facilitate piece-wise processing. First, the Blowfish algorithm was initialized, specifying the number of Blowfish rounds to be run. Next, user-defined S-box contents were initialized, and encryption transformations were applied to adjust the dimensions of ciphered layers. Finally, the encrypted image was formed by combining these layers and undergoing a bitwise XOR operation with the chaos based PRNG, resulting in a secure and uniquely encrypted image. The image dataset from Signal and Image Processing Institute (SIPI) database [78] was used for basic evaluations: correlation coeffecients, differential analysis attack and information entropy. There are no further evaluations such as data loss attack and noise attack analysis, or any comparison with traditional algorithms.

The implementation presented in [42] proposed an efficient image encryption algorithm that exhibited resilience against Single Event Upsets and transmission errors. It used AES-Counter (AES-CTR), as per the recommendation of onboard satellites [21], utilizing a chaotic block selector that operates on a newly introduced chaotic map - 2D Logistic-Adjusted-Sine (2D-LAS) map [97]. The encryption process started by deriving the keystream from the AES-CTR. Since a plain AES-CTR would treat an image as a binary stream, the image would end up encrypted, but without any distribution of the pixels or the keystream. To address this issue, a chaotic block selector relying on the 2D LAS chaotic map was utilized to choose a plain block from the image, and a key block from the keystream. The plain blocks were then XORed with the scrambled keystream, resulting in the final encrypted image. The study used Alsat-1, Alsat-2, and Deimos-2 satellite images with different resolutions and geographical information for evaluation. The algorithm also had a keyspace of $2^{284}$ to resist brute force attacks. Their evaluation showed that, if a Single Event Upset occurs during the keystream generation, only that specific block would get corrupted when decrypted. Also, if a transmission error occurred corrupting a single bit of a single pixel within the cipher block, only the corresponding pixel in the decrypted image gets corrupted due to AES-CTR usage. The implemented scheme proved high throughput with moderate power consumption, making it suitable in LEO satellites.

The satellite image encryption algorithm proposed in [65] adapted Fridrich's scheme [98] for multi-spectral satellite images. The confusion and diffusion processes of the algorithm were implemented separately, where the confusion was accomplished by using a 2D Cat map [99] for complex substitutions between the pixel positions. Diffusion was achieved by distributing the impact of each bit from the original image across the encrypted ones, inspired by the approach of [100]. The key generation was a modification to the proposed key generator in [100], where the discretized PieceWise Linear Chaotic Map (PWLCM) was used instead of the tent map. The key space was set to $2^{192}$, where the 192-bit was employed to establish the initial conditions and parameters for the PWLCM. To evaluate the algorithm, a unique Deimos-2 satellite image was utilized, captured over Vancouver, Canada, featuring a 1-meter resolution. This image had dimensions of (1311x873) and includes 4 spectral bands: near-infrared, red, green, and blue. The evaluation showed that the scheme is sensitive to Single Event Upsets, and therefore the proposed algorithm can only be used on LEO satellites, after incorporating Single Event Upsets mitigating solutions while increasing hardware complexity.

### 4.3. Enhanced Chaotic Maps

The third category we explored was the research work that enhanced chaotic maps for better and secure satellite image encryption process. Academia and industry have been keen on exploring innovative methods to secure the satellite image encryption process, specially by refining chaotic maps with different strategies such as expanding key spaces and enhancing encryption efficiency. This is necessary because of the recent advancements and wide usage of satellite communications, the number of cyber threat have been increased, making secure image transmission a challenge.

Authors of [69] proposed an enhanced chaotic map-based Isomorphic Elliptic Curve (IEC) for satellite image encryption, to address issues related to key distribution, small key sizes, and slow processing speeds. Elliptic Curve Cryptography (ECC) is a public-key cryptography technique, that provides a high-security with minor-key size compared to other cryptography techniques. The proposed approach has two main processes: first generating all the keys, and then constructing the key matrix to be used for encryption and decryption. A cryptosystem key generator has been used to generate all the keys in key schedule, while Elliptic Curve Diffie-Hellman (ECDH) has been used as the key exchange technique to ensure secure key distribution and key varying. The cascaded Modular Tent Logistic map generates a chaotic sequence (using the keys from IEC) to create a random number generator (RNG), which is later used to construct the key matrix for encryption and decryption processes. The cascaded Modular Tent Logistic map is a novel modular 1D chaotic system, which has been implemented by merging both Tent Map and Logistic map, to address the limitations of each individual map. Initially, the original image undergoes pre-processing, employing Discrete Wavelet Transform and subsequent partial Arnold Cat Map shuffling. Next, the image is encrypted using the Key Matrix. The first evaluation that was performed on three (512x512) satellite images showed that the proposed approach takes less time for encryption and decryption compared to [42], while boasting a larger key space of $2^{924}$. The second evaluation was performed on different image sizes, such as (1024x1024) and (512x512), to show the algorithm's robustness. The results confirmed that the algorithm is suitable for LEO satellites due to the high key space and speed, however, no analysis was performed for Single Event Upsets and transmission errors.

The encryption scheme proposed in [70], merged RNA computing [101] with a 7-Dimensional complex chaotic system [102]. The algorithm followed two steps: scrambling and diffusion. For the scrambling phase, the Arnold transform is integrated with chaotic sequences. An Arnold transform is an operation that rearranges the pixels in the image [103]. For the diffusion phase, the RNA codon (a triplet of three nucleotides in RNA) is disrupted by the Arnold transform. Finally, the pixel values of the scrambled image are diffused by employing the newly defined RNA codon and RNA operation rules, which are determined by the chaotic sequences. Scrambling and diffusion operations introduce unpredictability, disorder, and complexity. The paper utilized a color image of 'Lena' (256×256), and images 'Kyiv', 'San Diego', and 'Colosseum' (512×512), to conduct the evaluation of the proposed algorithm. A data loss attack was executed on the RGB image of San Diego, and the image quality significantly dropped when (256x256) blocks are removed from the image.

Authors of [71] used double tent maps to increase the key space [104] and to improve the confusion and diffusion within the satellite image encryption algorithm. Authors proposed a new Pseudo Random Number Generator (PNRG) that uses a double tent map to generate a sequence. With the use of two tent maps instead of one to increase the keyspace, the PRNG can resist brute-force attacks. Initially, in the confusion process, the position of the pixels is shuffled based on the sequence generated by PRNG. Then, the diffusion process is applied where the confused image is XORed with the PRNG generated key, producing the cipher image. Authors have performed the evaluation on a singular image, by conducting differential attacks and by computing correlation coefficients in horizontal, vertical, and diagonal directions. Additionally, the information entropy was calculated to quantify the level of randomness or certainty present in the data.

The satellite image encryption algorithm proposed in [72] improved the encryption schemes proposed in [105] and [106] by employing traditional AES with along chaotic maps. The authors used a two-dimensional Standard Chaotic map [107] and a simple form of diffusion in the process of confusion (mixing pixel positions), before encrypting the data using AES. Furthermore, a novel key generator was implemented based on Discrete Chaotic maps, namely Discretized Skew Tent Map (DSTM) and Discretized Logistic Map (DLM) to achieve a robust pseudo-random generator [108, 109]. The use of discretized versions of the original maps reduce the implementation complexity [100]. Also to increase resistance to brute force attacks, a larger key space of i is $2^{288}$ is used. The data sets used for evaluation consist of three panchromatic satellite images: Tripoli, Rio De Janeiro, and Stockholm (10 bits) from the WorldView-3 satellite [79], However, authors have not specified the size and resolution of these images. The authors performed a key sensitivity analysis to ascertain the impact of minor modifications in the secret key on the output produced by the cryptographic algorithm. Furthermore, correlation coefficients and information entropy are calculated to quantify the randomness of the encrypted image. Lastly, differential attacks were performed and NPCR (Number of Pixels Change Rate) and UACI (Unified Average Changing Intensity) values were calculated to see how well the algorithm holds to attacks. However, the authors have not performed further evaluations such as data loss attack and noise attack analysis, or any comparison with traditional algorithms.

## 4.4. Integration of Multiple Chaotic Maps

The fourth category we explored was the research work that integrated multiple chaotic maps for better and secure satellite image encryption process. As we discussed in previous sections, there are several work in literature that explored different types of chaotic maps, to develop secure and robust satellite image encryption algorithms. However, there are limited number of research that have been carried out to study the effect of combining different types of chaotic maps, with the objective of developing improved versions of satellite image encryption algorithms. Even though the integration of multiple chaotic maps would results in an algorithm that resist attacks, it is important to note that the overall complexity of the cryptographic process also increases.

The authors of [73] and [74], leverage the combined dynamics of chaotic maps to enhance encryption robustness. While the approach proposed in [73] combined Logistic Map, Gauss Iterated Map, and Henon Maps to generate a key matrix, the approach proposed in [74] integrates SHA-512 hashing and various chaotic maps such as Chebyshev, Tent, Cubic, Henon, Logistic and Sine maps.

The research work on [73], created a secret key using various chaotic maps such as Logistic Map, Gauss Iterated Map and Henon Map, to enhance the algorithm's confusion and diffusion processes. The study suggested a secret key matrix generation algorithm, where the key is initially generated using a Logistic Map, which is then used as an input to generate keys using the Gauss iterated map. Furthermore, another key is generated by employing the Henon map. The secret key matrix used for image encryption is created by XORing the three keys mentioned above. During the encryption phase, the image is segmented into blocks and then divided into its RGB components. The pixels of each component are then XORed with the corresponding Choas key matrix. In addition, exchange and recombination processes are applied to the sub-images. Finally, the three components are merged to obtain the encrypted image. The key space used in the algorithm was $2^{250}$ to resist brute force attacks. The study used satellite images obtained from the United States Geographical Survey for basic evaluation. The authors computed the correlation coefficient despite the absence of specific values for the test. Additionally, they calculated the information entropy, aligning with findings from other studies. Finally, differential attacks were executed to derive NPCR (Number of Pixels Change Rate) and UACI (Unified Average Changing Intensity) values which show how resistant the algorithm is to these kind of attacks. However, the authors have not performed further evaluations such as data loss attack and noise attack analysis, or any comparison with traditional algorithms.

The satellite image encryption algorithm proposed in [74], combined the SHA-512 function and chaos mapping parameters to provide high security and high sensitivity. The process has three steps, and the first step is converting a user-input key into a 512-bit hash code, using SHA-512 function. The hash code is further converted into a matrix of numbers using ASCII code. The key matrix obtained using ASCII code is divided into smaller matrices, on which statistical functions are applied. In the second step, Chebyshev, Tent, Cubic, Henon, Logistic, and Sine chaotic maps determine initial values and fixed parameters for chaos mapping. The values obtained from statistical functions and chaotic maps contribute to the formation of the encryption blocks for each map. As the third step, the satellite image and the encryption blocks are combined using XOR operations, and the resulting blocks are further combined with 512-bit hash values to create the final encryption block. Authors have conducted a basic evaluation by comparing the proposed approach with existing approaches to show the performances of the proposed approach. However, they have not performed further evaluations such as data loss attack and noise attack analysis.

## 4.5. Chaotic Oscillators and Controllers

The fifth category we explored was the research work that utilized chaos theory in designing oscillatory systems and control mechanisms, offering insights into application areas such as signal processing and chaos based secure communication. Authors of [75] introduced a secure transmission approach for satellite images with the use of chaotic oscillators. The proposed approach employed Chua chaotic oscillators [110] at the transmitter and receiver, which dynamically generate the key for encryption and decryption. Following Lyapunov stability theory and the finite-time synchronization concept, the chaotic controllers achieve finite time synchronization within a channel experiencing time delays. The proposed approach utilized multi-shift cipher encryption and chaotic masking to increase the security of the wireless Orthogonal Frequency Division Multiplexing (OFDM) wireless network. The authors aimed to remove the appearances of the original satellite image during the transmission, while preserving an adequate level of quality of the recovered satellite image. Experimental results have shown that the proposed scheme has a low time consumption and can be easily implemented in electronic circuits. The study used the Boston Satellite image (1400x700x3) for the evaluation, specifically for data loss attacks with (64x64) blocks and noise attacks involving salt and pepper.

The research work presented in [76] proposed a dual-channel key transmission method, where plaintext keys are sent using a distinctive bit-level key hiding strategy. This method was designed to withstand known-plaintext and chosen-plaintext attacks. During the image scrambling phase, a novel technique known as Cross-boolean select scrambling (CBSS) is introduced, which creates a boolean matrix through the use of the Chaotic Flow with the Plane of Equilibria (CFPE) system. Next, the boolean matrix is used for cross-mixing multiple bands in the image being transmitted. The proposed approach improves security and prevents the risk of the entire image being easily reconstructed, which is a vulnerability in traditional color image encryption single-channel methods. Additionally, the diffusion method is crafted by merging the semi-tensor product (a mathematical operation for combining two matrices) with the CFPE system.

## 5. Testing and Evaluation

Evaluating the performances of any encryption algorithm is essential to understand its effectiveness, robustness and suitability for practical applications. Assessing the security of chaos based encryption algorithms, ensures that they can withstand attacks. Furthermore, a rigorous evaluation helps security professionals to identify vulnerabilities and potential weaknesses of chaos based encryption approach and essentially provide guidance for improvements.

In the previous section, we have comprehensively discussed several research work that have explored chaos based image encryption approaches, specifically for satellite communication. These research work have conducted evaluations to prove the performances of the proposed approaches targeting different aspects such as security assurance, robustness, resistance to attacks and etc.

In this section, first, we discuss standard methods and parameters used for the evaluation of chaos based image encryption algorithms. Next, in Table 3, we provide a summary of the evaluations performed by the research work that we explored in the previous section, including different standard tests and cryptanalysis attacks.

- **National Institute of Standards and Technology (NIST) Encryption Test**:

  The NIST Cryptographic Algorithm Validation Program (CAVP), is a critical initiative for ensuring the security and reliability of encryption algorithms. Specifically, the NIST SP 800-22 Rev.1 provides a Statistical Test Suite for Random and Pseudorandom Number Generators used in cryptographic applications. It is a statistical software package that contains 16 different tests designed to evaluate the security and strength of an encryption algorithm. These tests include Frequency, Block Frequency, Runs, Spectral Analysis and Template Matching. They can test binary sequences of any length, generated by hardware or software based cryptographic random or pseudorandom generators. Successfully passing these tests signifies that the sequences generated by the algorithm possess genuine randomness.

- **Key Space**

  A key space is the set of all valid, possible, distinct keys of a given cryptosystem. Essentially, a larger key space increases the encryption's security, provided that the encryption algorithm is strong and properly executed. A larger key space offers a greater number of potential keys, complicating and prolonging the process for an attacker to identify the correct key. According to the literature, the size of the encryption algorithm's key space should exceed $2^{100}$ to avoid being prone to brute-force attacks [84].

  In chaos based encryption, the key space is often dynamically changeable. This adaptability enhances security by allowing a wide range of potential keys. An algorithm that associates the variable key space with the initial conditions of a chaotic map (such as the Hénon map), is an example. As the chaotic system evolves, the key space dynamically adjusts, making the crypto system extremely sensitive to the chosen key.

- **Key Sensitivity Analysis**:

  Key Sensitivity Analysis plays a crucial role in assessing the robustness and security of encryption algorithms [111]. It helps security analyst to understand how variations in encryption keys impact the resulting cipher text. Security analyst manipulate the encryption keys, introducing slight variations. The encryption algorithm processes the same plain text with these modified keys, producing distinct cipher texts. The differences between these cipher texts reveal the algorithm's sensitivity to key changes. Metrics used for assessment include: (1) Normalized Cross-Correlation Parameter (Cncc) that measures how closely related the ciphertexts are, (2) Pixel Correlation Coefficients that assess the correlation between adjacent pixels, (3) Number of Pixels Changing Rate (NPCR) that quantifies the impact of key variations and (4) Unified Averaged Changed Intensity (UACI) that evaluates overall changes in pixel values.

- **Correlation Coefficients Analysis**:

  Correlation Coefficients Analysis assesses the ability of the proposed encryption algorithm to resist statistical attacks. A good encryption algorithm should produce encrypted images where neighboring pixels exhibit low correlation, making it challenging for attackers to exploit patterns or relationships in the encrypted data. To conduct the analysis, first, several adjacent pixels are picked out vertically, horizontally, and diagonally [112]. Next, the correlation coefficient (CC) formula [113] is used to compute the correlation coefficient in all three directions. An effective image encryption algorithm is characterized by CC values approaching zero, indicating a stronger level of encryption as the correlation coefficient gets closer to zero.

  Chaos based encryption methods demonstrate low correlation coefficients, making them robust against statistical attacks.

- **Differential Attack**:

  A differential attack helps security analyst to understand how variations in input data impact the resulting cipher text. Security analyst first modifies pixels in a plain text image and subsequently encrypts both the original and the altered images. Next, he compares the differences between the two images to decipher them. By analyzing the differences in the resulting cipher texts, the security analyst gains insights into the key. The algorithm needs to be sensitive to any changes in the plain text image. The resistance of an image against differential attacks is quantified using: (1) Number of Pixels Changing Rate (NPCR), (2) Unified Averaged Changed Intensity (UACI) and (3) Hhamming Distance (HD) [114]. The optimal values for the NPCR, UACI, and HD of an encrypted image are 99.61%, 33.46%, and 50%, respectively. [115, 61].

  Chaos based schemes exhibit high sensitivity to initial conditions (pixel and key values). This sensitivity makes them resilient against differential attacks. As differential attacks involve analyzing the differences between plain text and cipher text, and chaos based encryption introduces sufficient randomness to thwart such attacks.

- **Information Entropy Analysis**:

  Information entropy analysis is used to measure the randomness and unpredictability of the encrypted data. In terms of an image, it shows the amount of information contained in each pixel. Higher entropy suggests greater unpredictability and complexity in the image data, indicating a stronger encryption algorithm [116, 117]. In a gray-scale image, as the information entropy approaches 8, the pixel values in encrypted image exhibit greater randomness.

  Noise attacks in satellite image transmissions can significantly impact the integrity and reliability of transmitted images. Attackers perform noise attacks on cipher images using different approaches such as Gaussian Noise, salt and pepper noise and speckle noise. As a result, the receiver may not be able to decrypt the image properly [118]. Therefore, it is important to evaluate the robustness of the encryption algorithms, which shows the ability of the encryption algorithm to withstand adversarial attempts, and produce valid cipher texts under different encryption keys. The quality of the decrypted image following a noise attack, can be assessed by the Peak Signal-to-Noise Ratio (PSNR) [119]. Additionally, the security analyst can perform an analysis for the data loss attack on the image encryption algorithm, which guarantees that, even if some data in the ciphered image is lost, the decrypted image should still preserve most of the original information [67]. Generally, a block of (32x32), (64x64) or (128x128) is removed from the cipher image, and PSNR is calculated to verify how much important information is retained.

- **Fault propagation analysis (Single Event Upsets (SEUs) and Transmission Errors)**:

  Satellites are exposed to high radiation, that cause faults in the computer systems deployed on the satellite. When encrypting images, a single-bit fault in the satellite's computers may propagate, causing multiple flaws in the cipher data, which is known as Single Event Upsets (SEUs) [120, 121].

  To check the impact of SEUs, two keys are utilized. First, one image is encrypted with the original first key, and next, the other image is encrypted with a keystream, in which one random bit is flipped. The NPCR and UACI values are calculated for both images and are compared [61]. In addition, images being transferred from the satellite to the ground, may be exposed to transmission errors. Two plain text images are encrypted using a secret key to test for transmission errors, and then a one-bit flip is incorporated into one of the images. The error-containing image is overlaid onto the first image randomly, creating a second encrypted image with distortions. The NPCR and UACI of the two versions are analyzed.

- **Histogram Analysis**:

  Histogram analysis is one of basic and compulsory evaluation for any image encryption algorithm, that assess the distribution of images. It provides insights into the distribution of pixel intensities within an image. The original image is expected to have an uneven distribution, whereas the encrypted image should exhibit a uniform distribution. A uniform distribution suggests a random image, implying a minimal likelihood of reconstructing the original image [122].

  Chaos based encryption schemes often yield highly scrambled encrypted images with uniform histogram distributions. This uniformity enhances security by minimizing patterns that could be exploited by attackers. Unlike conventional techniques, which may exhibit spikes or non-uniform distributions, chaos based schemes ensure a more balanced spread of pixel intensities.

**Table 3**
Comparative Analysis of Image Encryption Algorithms

| Research | Techniques | NIST Test | Key Space | Key Sensitivity | Correlation Coefficients (H, V, D) | Differential Attack (NPCR %, UACI %) | Information Entropy | Robustness Analysis (dB) | Fault Propagation |
|---|---|---|---|---|---|---|---|---|---|
| [61] 2023 | 2D filtering and Fisher Yates shuffling | ✓ | $2^{250}$ | ✓ | 0.0021, 0.0028, 0.0048 | 99.60, 33.45 | 7.9975 | 22.9896 dB | ✓ |
| [62] 2021 | Pseudo-random key combined Logistic Map, Cosine Transformed Logistic Map and Cosine Transformed Logistic-Sine Map (CTLSM) | ✗ | $2^{384}$ | ✓ | 0.0005, 0.0009, 0.00043 | 99.63, 33.44 | 7.9998 | ✓* | ✗ |
| [63] 2019 | Chaotic Logistic Map with crossover, mutation and scrambler operations | ✓ | $2^{398}$ | ✓ | -0.0016, 0.0010, -0.0017 | 99.6414, 33.4265 | 7.9970 | ✗ | ✗ |
| [64] 2017 | Logistic, Sine and Standard Maps with cyclic rotation based on Graph Theory | ✗ | $2^{512}$ | ✓ | ✓* | 99.60, 34.60 | 7.9963 | ✗ | ✗ |
| [42] 2020 | 2D Logistic Adjusted Sine Map | ✗ | $2^{284}$ | ✓ | 0.0004, 0.0005, 0.0003 | 99.61, 33.48 | 7.9994 | ✗ | ✗ |
| [65] 2020 | Adaptation of the Fridrich's Scheme for multi spectral satellite images and Uses 2D Cat Map for confusion | ✗ | $2^{192}$ | ✓ | 0.042, -0.038, -0.012 | 99.60, 33.47 | ✗ | ✗ | ✗ |
| [66] 2023 | 7D Hyperchaotic System, A pseudo-AES algorithm and An arrangement permutation | ✓ | $2^{350}$ | ✓ | 0.0052, -0.0031, -0.0004 | 99.62, 33.44 | 7.9976 | 20.1120 | ✓ |
| [67] 2021 | Linear Feedback Shift Register (LFSR) generator, SHA-512 hash function, Hyperchaotic systems, and Josephus problem | ✓ | $2^{358}$ | ✓ | 0.0018, -0.0020, -0.0012 | 99.62, 33.45 | 7.9977 | 28.1675 dB | ✓ |
| [68] 2021 | 3D Henon and Chen Chaotic Maps, with Blowfish encryption | ✓ | $2^{448}$ | ✗ | -0.0063, -0.0038, 0.0017 | 99.6098, 32.9168 | 7.9994 | ✗ | ✗ |
| [69] 2021 | Cascade Modular Tent Logistic Map (CMTLM) and Isomorphic Elliptic Curve Cryptography (IECC) to increase the key-space | ✗ | $2^{924}$ | ✓ | 0.0002, 0.0004, 0.0001 | 99.61, 33.46 | ✗ | ✗ | ✗ |
| [70] 2023 | RNA and 7D Complex Chaotic system | ✓ | ✗ | ✓ | -0.0006, -0.0021, -0.0005 | 99.61, 33.43 | 7.9985 | 25.1331 | ✓ |
| [71] 2018 | PRNG using Double Tent Maps | ✗ | $2^{208}$ | ✗ | ✗ | ✗ | 7.9971 | ✗ | ✗ |
| [72] 2020 | AES with Discretised Chaotic Maps | ✗ | $2^{288}$ | ✓ | 0.0016, 0.002, 0.002 | 99.89, 33.36 | 9.9996 | ✗ | ✗ |
| [73] 2018 | Logistic and Henon Maps | ✗ | $2^{250}$ | ✓ | ✓* | 0.9965, 0.3347 | 7.9973 | ✗ | ✗ |
| [74] 2019 | Chebyshev, Tent, Cubic, Henon, Logistic and Sine Maps with SHA-512 | ✗ | ✗ | ✗ | ✗ | 99.61, 33.50 | 7.9992 | ✗ | ✗ |
| [75] 2021 | Finite Time Chaos S approach | ✗ | $2^{419}$ | ✓ | ✓* | 99.6159, 34.3190 | 7.9986 | ✓* | ✗ |
| [76] 2021 | Cross-Boolean Select Scrambling (CBSS) using Chaotic Fow with Plane of Equilibrias (CFPE) system | ✗ | $2^{585}$ | ✗ | 0.0007, 0.0064, 0.0004 | 99.61, 33.47 | 7.9993 | ✓* | ✗ |
| [123] 2017 | AES-CTR with GEFFE generator | ✗ | ✗ | ✗ | -0.0016, -0.0018, -0.0024 | ✗ | ✗ | ✗ | ✓ |

*Asterisk indicates that the value was not provided in the source.

## 6. Challenges and Future Research Directions

Chaos based image encryption is considered to be an ideal approach for satellite communication, because of its high flexibility, less computational overheads, less computing power and ease of implementation. However, with the increasing number of cyber threats targeting satellite communication systems, it is crucial to deploy secure encryption schemes that address existing vulnerabilities as well as anticipated future threats. In this section, we will explore the current challenges faced by chaos based image encryption, and possible future research directions to improve the existing chaos based image encryption schemes.

### 6.1. Defending Zero Day Attacks

Even though the chaos based image encryption is considered to be an ideal approach for satellite communication, there is no guarantee that the chaos based encryption scheme is 100% secured. Chaos based encryption algorithms often suffer from inherent instability due to periodic behaviors in their mappings, and these periodicities can be exploited by attackers, compromising system security. Finite computational precision further exacerbates this vulnerability. The study by [124] notes that any chaos based encryption algorithm that uses a constant key is vulnerable, and plain text can be easily retrieved.

Zero day attacks present significant challenges for chaos based image encryption. Zero day attacks can expose unknown vulnerabilities of chaos based encryption algorithms at any given time, and they can be very problematic for satellite communications due to the critical nature. Zero day attacks emerge without warning, catching defenders off guard. Traditional security measures that rely on historical data are inadequate for handling novel zero day threats. Furthermore, zero-day exploits often involve multiple stages, and employ emerging technologies such as Machine Learning, Neural Networks and Deep Learning to deploy sophisticated attacks [125]. Zero-day attacks underscore the intricate nature of chaos based encryption, urging researchers to develop robust methods while acknowledging the inherent difficulties in chaotic systems. Therefore, it is a timely requirement to explore novel chaos based image encryption algorithms to ensure secure satellite transmissions.

### 6.2. Comprehensive Cryptanalysis

Due to the inherently uncontrollable nature of the trajectories, chaos based encryption algorithms present unique challenges in cryptanalysis [126]. First, chaotic systems are highly sensitive to initial values and parameters. Even minor perturbations can lead to vastly different trajectories. Cryptanalysts attempting to break chaos based encryption must consider with sensitivity which makes prediction and analysis complicated. Second, chaotic systems generate pseudo-random sequences, challenging the predictability. Unlike traditional cryptography algorithms with well-defined rules, chaos based encryption introduces an element of randomness that complicates the cryptanalysis. Deciphering these sequences requires specialized techniques and a deep understanding of chaotic behavior. Third, chaos based encryption relies on mappings derived from chaotic systems. Analyzing these mappings to identify weaknesses is not a a straightforward process. The complex interplay of parameters, nonlinearity, and topological transitivity, demands sophisticated cryptanalysis methods.

Due to these challenges, chaos based encryption algorithms have been considerably less subject to cryptanalysis than traditional encryption algorithms such as AES, DES and 3DES. Cryptanalysts must ensure that chaos based encryption remains resilient even when subjected to aggressive scrutiny. There are very limited work has been done on cryptanalysis of chaos based encryption algorithms to reveal their weaknesses [127]. Chaos based encryption algorithms require novel comprehensive cryptanalysis approaches, that can prove algorithm's resistance to wide range of attacks. Therefore, it is crucial to explore novel cryptanalysis approaches, so that the weaknesses of the chaos based encryption can be identified and mitigation strategies can be deployed at the early stages.

### 6.3. Quantum Threats

Quantum computing, with its superior computational capabilities, poses a significant threat to classical encryption schemes, including chaos based encryption [128]. Theoretically, quantum computers can solve certain problems much faster than classical computers potentially breaking many current encryption algorithms. The emergence of quantum computers has created a new challenge for existing security algorithms, commonly called the *quantum threat* [128]. Quantum computers can potentially break public-key cryptography systems such as RSA, and current public-key cryptography is expected to be broken by a large-scale quantum computer in future. Fortunately, according to the existing research work, chaos based encryption is still resistant to quantum attacks. The study [129] proposed a four-dimensional chaos-based quantum image encryption technique1. This algorithm addresses the shortcomings of Arnold transformation periodicity, small key space, and the lack of resistance to statistical analysis, proposing a reliable and effective encryption scheme1. Authors of [130] designed a novel encryption algorithm for quantum images based on chaotic maps. The image is converted into a scrambled state using quantum circuits, and the scrambled image is encrypted using the quantum XOR gate with the chaotic maps algorithm.

Even though chaos based encryption is still resistant to quantum attacks, quantum computing is a rapidly evolving field, and therefore, it is necessary to explore novel chaos based encryption methods that cannot be broken by quantum computers in future. Cryptographers are designing new algorithms to prepare for Q-Day, the day when current algorithms will be vulnerable to quantum computing attacks *Post quantum cryptography* referred to as quantum-proof, quantum-safe, or quantum-resistant, aims to develop cryptographic algorithms (usually public-key algorithms) that are supposed to be secure against a cryptanalytic attack by a quantum computer [128]. Furthermore, *Quantum Agile* in encryption refers to the need for cryptographic systems to be agile and adaptable in the face of the potential threats posed by quantum computing [128]. The cryptographic systems should have the crypto-agility, which is the ability to quickly and securely change cryptographic algorithms and related data, in the case of their compromise. while quantum computing presents significant challenges to current cryptographic systems, it also opens up new avenues for secure communication. The field is rapidly evolving, and it is crucial for researchers to actively work on developing quantum-resistant and quantum agile chaos based encryption methods.

### 6.4. Hardware Implementations

While chaos based encryption proves to be a promising approach for satellite image encryption, practical implementation requires comprehensive considerations [131]. Since most of the proposed approaches require a specific hardware implementation, or integration with existing systems, ensuring compatibility and interoperability is crucial. As hardware platforms such as Field Programmable Gate Arrays (FPGAs) have limited resources, designing efficient and compact chaos based encryption circuits while meeting performance requirements is complicated [132, 133]. The deployment of chaos based algorithms either a new approach or any modification to existing approach, is currently hampered by prohibitive costs and hardware requirements. Therefore, it is important for the researches to explore cost effective and less complicated hardware implementations for chaos based encryption systems, while addressing resource constraints, speed-security trade-offs, and compatibility with existing systems.

### 6.5. Integrating chaos based encryption with traditional encryption methods

Chaos based encryption brings unpredictability and pseudo-randomness due to chaotic dynamics. When integrated with traditional encryption algorithms such as AES or RSA, chaos based encryption can augment the security properties. The combination leverages the strengths of both chaos and conventional encryption, enhancing security, robustness, and efficiency. It would be interesting to explore hybrid encryption schemes that blend chaos based algorithms with symmetric or asymmetric encryption. As an example, *Chaos-Stream Cipher Hybrid*, where the chaotic maps generate a stream of keys, which are then XORed with plaintext [134]. This approach combines chaos's unpredictability with the efficiency of stream ciphers. Another example is *Chaos-Block Cipher Hybrid*, where chaotic sequences modify the key schedule or substitution boxes in block ciphers [134]. The resulting hybrid cipher balances security and speed. Therefore, it is interesting to explore integrating chaos based encryption with traditional methods which opens a more resilient and versatile encryption landscape.

## 7. Conclusion

The rapid adoption of satellite communication in different applications has widened the threat surface of satellite systems, in-terms of physical security as well as cybersecurity. Satellite communications are susceptible to many cybersecurity vulnerabilities, such as hacking, malware infections, data breaches and supply chain attacks. Incorporating robust encryption and authentication protocols is essential for securing satellite communication systems, where advanced encryption algorithms are expected to protect the confidentiality and integrity of data transmitted via satellites, while strong authentication mechanisms are expected to prevent unauthorized access. Widely used services with satellite communication such as climate change monitoring, military surveillance and real-time event broadcasting, involve data in the form of images rather text. Therefore, even though the widely adopted method for satellite image encryption is Advanced Encryption Standard (AES), the nature of image data requires the exploration of encryption algorithms distinct from the conventional algorithms.

In this paper, we specifically focus on chaos based image encryption, and its usage for satellite communication, because of the advantages provided by chaos based image encryption, such as high flexibility, high security, less computational overheads, less computing power and ease of implementation. To the best of our knowledge, this is the first survey that provides a structured overview of chaos based image encryption, specifically focusing on Satellite images and Satellite communication. We started by presenting background knowledge on satellite communication and image encryption in satellite communication, directing our study to cover theoretical aspects of chaotic systems and its practical usage for image encryption. Highlight of our study was the comprehensive literature review on all state-of-the-art studies focusing specifically the chaos based satellite image encryption, which was further strengthen by the detailed analysis of the evaluation process, including evaluation parameters and conditions. Finally we discussed the existing challenges and open research problems for chaos based satellite image encryption, to facilitate future directions to upcoming research initiatives.

## Declaration of competing interest

The authors declare that they have no known competing financial interests or personal relationships that could have appeared to influence the work reported in this paper.

## Data availability

No data was used for the research described in the article.

## CRediT authorship contribution statement

**Farrukh Bin Rashid**: Conceptualization, Methodology, Investigation, Resources, Writing - Original. **Windhya Rankothge**: Conceptualization, Writing - Original, Supervision. **Somayeh Sadeghi**: Conceptualization, Writing - Review. **Hesamodin Mohammadian**: Conceptualization, Writing - Review. **Ali Ghorbani**: Supervision.